\documentclass[reprint, superscriptaddress, secnumarabic, amssymb, nobibnotes, aps, prl]{revtex4-1}

\usepackage{graphicx}
\usepackage{epstopdf}
\usepackage[T1]{fontenc}
\usepackage[latin9]{inputenc}
\usepackage{amsbsy}
\usepackage{gensymb}
\usepackage{textcomp}
\usepackage[T1]{fontenc}
\usepackage[latin9]{inputenc}
\usepackage{amsmath}
\usepackage{amssymb}
\usepackage{bbm}
\usepackage{soul}
\usepackage{physics}
\usepackage{braket}
\usepackage{xcolor}
\allowdisplaybreaks
\usepackage{graphicx}
\usepackage[colorlinks=true]{hyperref}  
\hypersetup{
    unicode=false,          
    pdftoolbar=true,        
    pdfmenubar=true,        
    pdffitwindow=false,     
    pdfstartview={FitH},    
    pdftitle={Microscopic insights into the superconducting properties of an equiatomic hexagonal high entropy alloy revealed by muon spectroscopy{}},    
    pdfauthor={, , ,},     
    pdfsubject={},   
    pdfcreator={},   
    pdfproducer={},  
    pdfkeywords={} {} {}, 
    pdfnewwindow=true,      
    colorlinks=true,       
    linkcolor=blue, 
    citecolor=blue,        
    filecolor=magenta,      
    urlcolor=blue           
}
\usepackage[normalem]{ulem}

\newcommand{\figref}[1]{Fig.~\ref{#1}}

\begin{document}
\title{Unveiling superconducting properties of an equiatomic hexagonal high entropy alloy via muon spin relaxation and rotation measurement}

\author{Sonika Jangid}
\affiliation{Department of Physics, Indian Institute of Science Education and Research Bhopal, Bhopal, 462066, India}

\author{Pavan Kumar Meena}
\affiliation{Department of Physics, Indian Institute of Science Education and Research Bhopal, Bhopal, 462066, India}

\author{Niraj P. Atale}
\affiliation{Department of Physics, Indian Institute of Science Education and Research Bhopal, Bhopal, 462066, India}

\author{Rhea Stewart}
\affiliation{ISIS Facility, STFC Rutherford Appleton Laboratory, Didcot OX11 0QX, United Kingdom}

\author{Adrian D. Hillier}
\affiliation{ISIS Facility, STFC Rutherford Appleton Laboratory, Didcot OX11 0QX, United Kingdom}

\author{R. P. Singh}
\email[]{rpsingh@iiserb.ac.in}
\affiliation{Department of Physics, Indian Institute of Science Education and Research Bhopal, Bhopal, 462066, India}

\begin{abstract}
Superconducting high-entropy alloys (HEAs) present a unique platform for studying the effect of disorder, composition, and crystal structure on superconducting pairing. In this study, we present a comprehensive bulk and microscopic investigation of the rarely observed equiatomic hexagonal HEA Nb-Mo-Ru-Re-Ir using magnetization, resistivity, heat capacity, and muon spin relaxation and rotation ($\mu$SR) measurements. Our findings confirm bulk type-II superconductivity with a transition temperature of 4.63(2) K and a high upper critical field. Heat capacity and transverse-field $\mu$SR data reveal conventional s-wave superconductivity, while zero-field $\mu$SR results suggest the preservation of time-reversal symmetry in the superconducting state. These findings provide valuable insights into the superconducting pairing mechanism in disordered multicomponent systems.
\end{abstract}
\maketitle

\section{Introduction}

High-entropy alloys (HEAs) represent a paradigm shift in materials science, combining five or more principal elements in near-equimolar ratios to create systems with exceptional mechanical properties and unique electronic behaviors \cite{yeh2004nanostructured, jien2006recent, gludovatz2014fracture, zou2015ultrastrong, lee2007effect}. The recent discovery of superconducting HEAs, a novel combination of exceptional mechanical properties of HEA and the phenomenon of superconductivity, has opened exciting new frontiers in both fundamental research and practical applications \cite{kovzelj2014discovery}. These HEA superconductors represent a novel class of disordered superconductors exhibiting the exotic phenomenon of superconductivity. They show remarkable characteristics, including superconductivity at extreme pressures, high critical fields and current densities, varied Debye temperatures, and unusual heat capacity behaviors \cite{robustSCpressure, motla2023superconducting, jangid2024superconductivity, jc, heathinfilm, kasem2021anomalous, marik2019superconductivity, motla2021probing}. The vast compositional landscape of HEAs allows for precise tuning of superconducting properties,  providing a unique opportunity to tailor superconducting behavior in ways that are difficult, if not impossible, to achieve through conventional methods like doping or high pressure. This ability to precisely engineer superconducting properties through HEA design significantly enhances the viability of applications in demanding fields such as aerospace, nuclear fusion, and high-field magnet technologies \cite{funct, dada2023recent, radiation, jangid2024superconductivity, jangid2025high}.

Despite a decade of research, the fundamental nature of superconducting pairing mechanisms in high-entropy alloys (HEAs) remains elusive. The frequent observation of conventional superconductivity in these systems is particularly perplexing, given their inherent high disorder typically detrimental to Cooper pair formation and the apparent absence of conventional phonon modes associated with BCS theory \cite{motla2022boron, motla2022superconducting, motla2023superconducting, jangid2024superconductivity, heathinfilm, status_cava, kitagawa2020cutting}. To elucidate the exact pairing mechanisms in HEAs, further investigation is essential. The critical limitations are that most HEAs, including superconducting HEAs, crystallize in FCC and BCC structures, and limited microscopic studies on superconducting HEAs. Therefore, detailed superconducting properties using bulk measurements along with microscopic studies of HEAs with different crystal structures are crucial to advance our understanding and shed light on these intriguing superconducting phenomena.

Hexagonal high-entropy alloys (HEAs), particularly those exhibiting superconductivity, are exceptionally rare, and the majority are stable only under high pressure. Their robust hardness and emerging application in superconducting devices for extreme conditions highlight their significant potential \cite{marik2019superconductivity, qiao2018rare, feuerbacher2015hexagonal}. In this paper, we present the synthesis and characterization of hexagonal superconducting high-entropy alloy (HEA) Nb-Mo-Ru-Re-Ir at ambient pressure by combining 4d and 5d elements in an equiatomic ratio to maximize configurational disorder and mitigate any dominant element-specific superconducting contributions. Notably, the inclusion of high-Z elements also provides a unique opportunity to investigate the role of spin-orbit coupling (SOC) in the superconducting ground state of these disordered alloys. The detailed investigation of the normal and superconducting properties of Nb-Mo-Ru-Re-Ir was carried out using magnetization, electrical and thermal transport, and muon spin relaxation and rotation ($\mu$SR) measurements. Our findings demonstrate that Nb-Mo-Ru-Re-Ir is a type-II superconductor with a transition temperature of 4.63(3) K, exhibiting conventional BCS superconductivity. Furthermore, $\mu$SR investigations confirm the preservation of the time-reversal symmetry in the superconducting ground state.

\begin{figure*}
\includegraphics[width=2.0\columnwidth, origin=b]{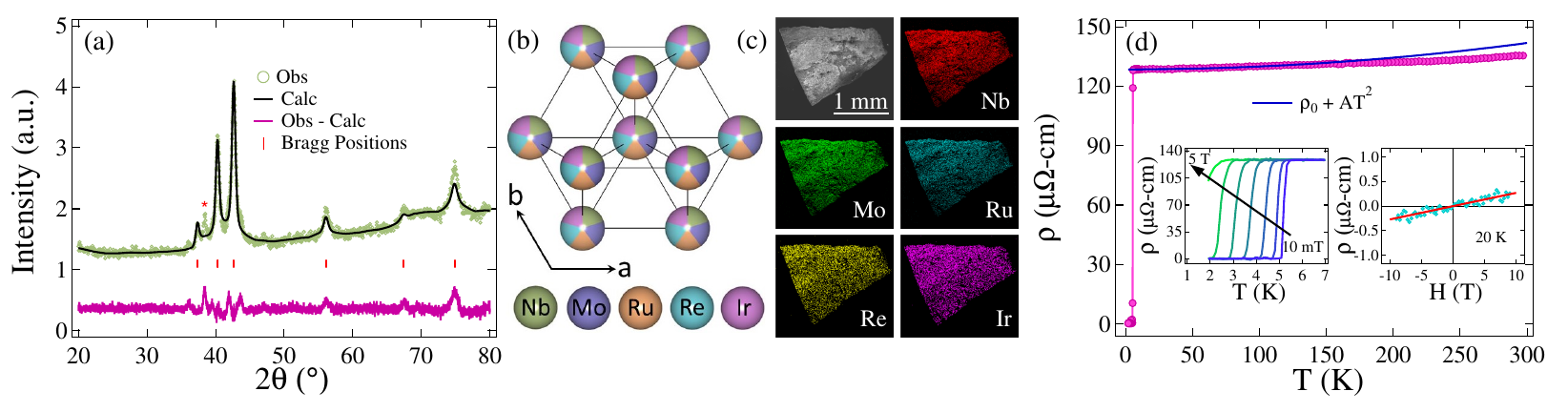}
\caption{\label{Fig1}(a) Room temperature powder X-ray diffraction pattern of Nb-Mo-Ru-Re-Ir, confirming the crystallization in a hcp structure. (b) Crystal structure of hcp Nb-Mo-Ru-Re-Ir with high site mixing. (c) EDX elemental mapping of Nb, Mo, Ru, Re, and Ir elements. (d) Electrical resistivity from base temperature (1.9 K) to room temperature under zero field. The left inset shows resistivity vs temperature curves for a range of applied fields, and the right inset shows the Hall resistivity from -9 T to 9 T at 20 K. }
\end{figure*}

\vspace{-1em}
\section{Experimental Details}
\vspace{-1em}
Polycrystalline Nb$_{0.20}$Mo$_{0.20}$Ru$_{0.20}$Re$_{0.20}$Ir$_{0.20}$ was synthesized via arc melting of high purity (4N) constituent elements in the stoichiometric ratio in an argon environment. To remove any residual oxygen, a titanium getter was used. To ensure homogeneity, the ingot was flipped over and remelted multiple times. The resulting sample exhibited significant hardness and negligible weight loss.

The phase composition was determined using an energy-dispersive X-ray analysis (EDAX). The identification of the phase was performed by powder X-ray diffraction using a PANalytic X$^{'}$Pert diffractometer with CuK$_{\alpha}$ ($\lambda = 1.5406 \text{\AA}$) radiation. Magnetization measurements were conducted using a Quantum Design Magnetic Property Measurement System (MPMS 3). The electrical resistivity and heat capacity measurements were conducted using a Quantum Design Physical Property Measurement System (PPMS), utilizing a four-probe and two-tau time-relaxation technique. The muon spin relaxation and rotation measurements were conducted using the MuSR spectrometer in both transverse and longitudinal configurations, which uses spin-polarized muons from the ISIS Neutron and Muon source at the Rutherford Appleton Laboratory in the United Kingdom \cite{hillier2019muons}.

\section{Results and Discussion}

\subsection{Sample Characterization}
\figref{Fig1}(a) presents the powder X-ray diffraction (XRD) pattern of Nb-Mo-Ru-Re-Ir. Le Bail refinement, performed using HighScore Plus software \cite{le1988ab}, confirms a hexagonal close-packed (hcp) lattice with space group P6$_3$/mmc (194). The refined lattice parameters are $a=b=2.77(8)$ \text{\AA} and $c=4.47(5)$ \text{\AA}. The observed broadening of the XRD peaks indicates a high degree of disorder \cite{kovzelj2014discovery}. A minor unindexed peak, corresponding to unreacted Ru and marked by a red asterisk, is detected; however, it does not affect the superconducting properties.

\figref{Fig1}(b) illustrates the atomic arrangement within the hcp structure of Nb-Mo-Ru-Re-Ir, where all elements can occupy the same Wyckoff position, resulting in significant site mixing. Energy-dispersive X-ray (EDAX) analysis at various sample locations determined an average composition of Nb$_{0.22}$Mo$_{0.21}$Ru$_{0.21}$Re$_{0.18}$Ir$_{0.18}$, which aligns with the nominal composition within experimental uncertainty. Furthermore, elemental mapping via EDAX confirms a uniform distribution of Nb, Mo, Ru, Re, and Ir, verifying the homogeneity of the phase, as shown in \figref{Fig1}(c).

\vspace{-1em}
\subsection{Normal and Superconducting Properties}
\begin{figure*}[ht]
\includegraphics[width=2.0\columnwidth, origin=b]{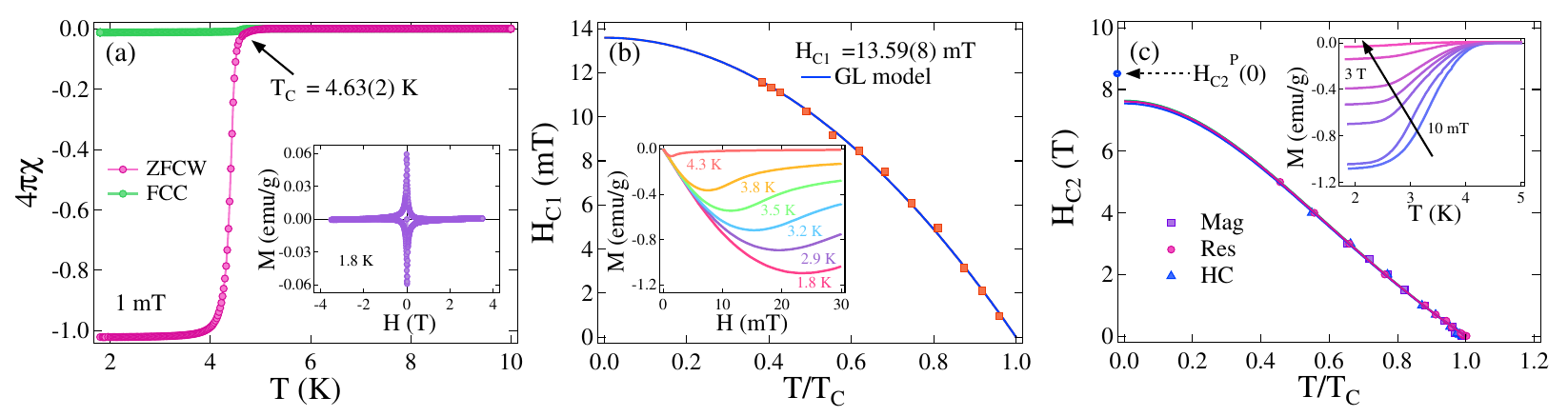}
\caption{\label{Fig2}(a) The temperature-dependent magnetic susceptibility in ZFCW and FCC modes under an applied field of 1 mT. The inset shows the magnetization loop at 1.8 K, confirming type-II superconductivity in Nb-Mo-Ru-Re-Ir. (b) Temperature-dependent lower critical field $H_{C1}(T)$ where the solid blue line is the fit to Eq. \ref{eqn2:Hc1}. The inset shows magnetization vs field curves taken at different temperatures. (c) Temperature-dependent upper critical field $H_{C2}(T)$ from magnetization (purple squares), resistivity (pink circles), and heat capacity (blue triangles) measurements fitted with the GL model. The inset shows magnetization vs temperature curves taken at different applied fields.}
\end{figure*}
\textbf{Electrical resistivity:}
Zero field resistivity measurements on Nb-Mo-Ru-Re-Ir, conducted from 1.9 to 300 K (see \figref{Fig1}(d)) show a sharp superconducting transition at $T_C = 5.23(3)$ K. 
Residual resistivity ratio (RRR) is calculated as $\rho_{300 K}/\rho_{10 K}=1.05(2)$, which, being close to unity, reflects the poor metallic behavior and the pronounced disorder inherent to HEAs \cite{motla2021probing, motla2022boron, motla2022superconducting, motla2023superconducting, jangid2024superconductivity}.
The degree of electron-electron correlation can be quantified using the Kadowaki-Woods ratio, defined as $K_w = A/\gamma^2_n$ \cite{kadowaki1986universal}. The coefficient $A$ corresponds to the contribution of electron-electron scattering to electronic resistivity at low temperatures, while $\gamma_n$ is the Sommerfeld coefficient derived from heat capacity measurements. $A$ is determined by the power law ($\rho=\rho_0+AT^2$) fitting of normal state resistivity at low temperatures (\figref{Fig1}(d)). The resultant value of $K_w$ is $10.3(6)\ \mu\Omega$-cmK$^{2}$J$^{-2}$mol$^{2}$
by utilizing the value of $A=1.48(3)\times10^{-4}\ \mu\Omega$-cmK$^{-2}$ and $\gamma_n=3.78(9)$ mJ-mol$^{-1}$K$^{-2}$. This value of $K_w$ is close to $10\ \mu\Omega$-cmK$^{2}$J$^{-2}$mol$^{2}$ suggesting a significant degree of electron-electron correlation in this system similar to heavy fermionic systems \cite{miyake1989relation, takimoto1996relationship} which has also been observed in other HEAs \cite{kim2020strongly, jangid2025high}. Furthermore, resistivity measurements at various applied magnetic fields were also carried out (shown in the left inset of \figref{Fig1}(d)) to determine the upper critical field, which will be discussed in the subsequent section. Furthermore, carrier concentration was determined by the field-dependent ($\pm$9T) resistivity $\rho_{xy}(H)$ at 20 K (in normal state), as shown in the right inset of \figref{Fig1}(d). 
A linear fit to the measured data gave a Hall coefficient $R_\mathrm{H} = 2.6(5)\times10^{-10}\ \ohm$ -mT$^{-1}$, revealing that holes are the dominant charge carriers. Applying the relationship $R_\mathrm{H} = 1/ne$, we obtained the carrier concentration $n=2.4(5)\times10^{28}\ \mathrm{m}^{-3}$.

\textbf{Magnetization:} 
\figref{Fig2}(a) shows the temperature variation of magnetic susceptibility in zero-field cooled-warming (ZFCW) and field-cooled-cooling (FCC) modes with an applied magnetic field of 1.0 mT. The transition from the normal to the superconducting state is marked by a diamagnetic response, with an onset at 4.63(2) K, corresponding to T$_C$. The bifurcation between the ZFCW and FCC modes below $T_C$ is attributed to strong flux-pinning. The isothermal magnetization loop at 1.8 K, as demonstrated in the inset of \figref{Fig2} (a), reveals type II superconductivity. Furthermore, isothermal magnetization with low field variation was also measured from 1.8 K to 4.5 K to determine the lower critical field (inset of \figref{Fig2}(b)). The lower critical field for each isotherm $H_{C1}(T)$ was determined as the field value at which the magnetization deviates from linearity. The temperature evolution of the lower critical field ($H_{C1}$) is best described by the Ginzburg-Landau (GL) equation as
\begin{equation}
    H_{C1}(T)=H_{C1}(0)\left[1-\left(\frac{T}{T_{C}}\right)^{2}\right]
    \label{eqn2:Hc1} 
\end{equation}
\begin{figure*}[ht] 
\includegraphics[width=2.0\columnwidth, origin=b]{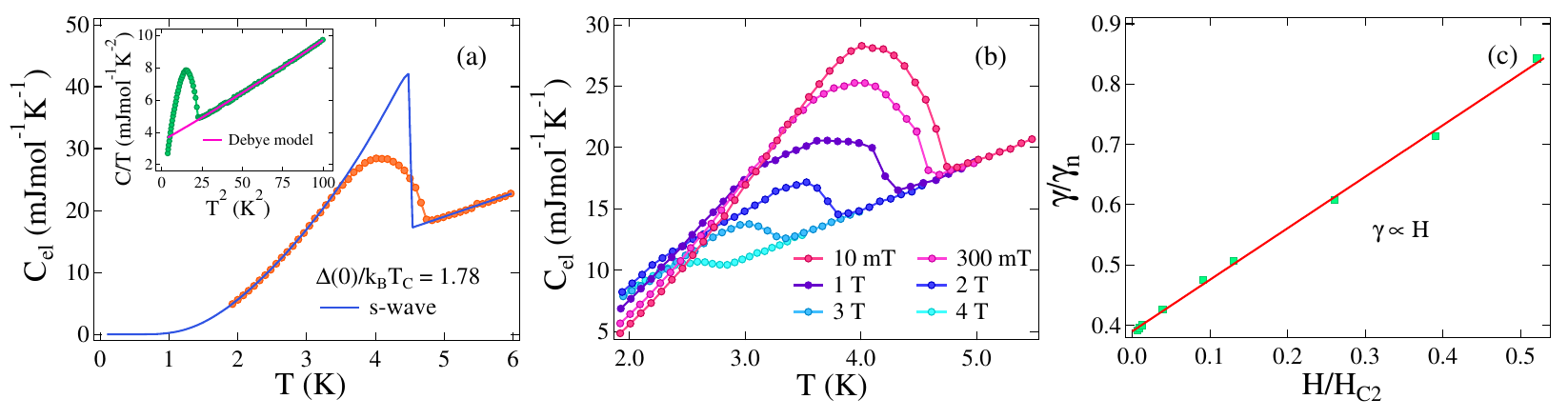}
\caption{\label{Fig3} (a) The temperature evolution of the electronic heat capacity under zero magnetic field fitted with Eq. \ref{eqn11:s-wave}. The inset shows $C/T$ versus $T^2$ in a zero applied field where the solid pink line is fit to Eq. \ref{eqn7:Debye}. (b) Temperature-dependent electronic heat capacity at various applied fields. (c) The field variation of $\gamma$.  }
\end{figure*}Fitting $H_{C1}(T)$ with Eq. \ref{eqn2:Hc1} results in an extrapolated value of $H_{C1}(0) = 13.59(8)$ mT as shown in \figref{Fig2}(b). Additionally, temperature-dependent magnetization, resistivity, and heat capacity measurements were conducted at different fields to determine the upper critical field ($H_{C2}$), where the magnetization vs temperature measurements are shown in the inset of \figref{Fig2}(c). With increasing the applied field, the superconducting transition temperature, $T_C$, shows a systematic decrease, and the upper critical field, $H_{C2}$, was extracted from this field-dependent behavior. The following GL equation is used to fit the temperature dependence of $H_{C2}(T)$, as shown in \figref{Fig2}(c). 
\begin{equation}
H_{C2}(T) = H_{C2}(0)\left[\frac{(1-t^{2})}{(1+t^2)}\right]
\label{eqn3:Hc2}
\end{equation}
where $t=\frac{T}{T_C}$ denotes the reduced temperature. By extrapolating the fit to 0 K, $H_{C2}(0)$ is determined to be 8.0(2), 7.61(3), and 7.68(8) T from the magnetization, resistivity, and heat capacity measurements, respectively.
$H_{C2}$ is directly related to the GL coherence length, $\xi_{\mathrm{GL}}(0)$ as follows \cite{Tinkham}:
\begin{equation}
    H_{C2}(0) = \frac{\phi_0}{2\pi \xi^2_{\mathrm{GL}}(0)}
    \label{eqn4:xi_GL}
\end{equation}
where $\phi_0 = 2.07\times10^{-15}$ T-m$^2$ represents the magnetic flux quantum. The GL coherence length is calculated to be $\xi_{\mathrm{GL}}(0) = 6.4(1)$ nm using $H_{C2}(0)=8.0(2)\ \mathrm{T}$. The other characteristic length, penetration depth $\lambda_{\mathrm{GL}}(0)$, can be determined with the help of the following equation which connects $H_{C1}(0)$, $\xi_{\mathrm{GL}}(0)$ and $\lambda_{\mathrm{GL}}(0)$ \cite{klimczuk2007physical}:
\begin{equation}
    H_{C1}(0) = \frac{\phi_0}{4\pi \lambda^2_{\mathrm{GL}}(0)}\left(\ln{\frac{\lambda_{\mathrm{GL}}(0)}{\xi_{\mathrm{GL}}(0)}+0.12}\right)
    \label{eqn5:lambda_GL}
\end{equation}
$\lambda_{\mathrm{GL}}(0)$ is determined to be 208(9) nm after replacing the values of $H_{C1}(0)$ and $\xi_{\mathrm{GL}}(0)$ in Eq. \ref{eqn5:lambda_GL}. The classification of superconductors into type I and type II can be done using the GL parameter ($\kappa_{\mathrm{GL}}$), which is the ratio of penetration depth and coherence length, $\kappa_{\mathrm{GL}}=\frac{\lambda_{\mathrm{GL}}(0)}{\xi_{\mathrm{GL}}(0)}$. Inserting the values mentioned above, $\kappa_{\mathrm{GL}}$ is determined to be 32.5(6), which is significantly larger than $1/\sqrt{2}$, further confirming type-II superconductivity in Nb-Mo-Ru-Re-Ir. 

An external magnetic field greater than the upper critical field can destroy the superconductivity. For a type-II superconductor, this can be accomplished via two processes: the orbital limiting effect and the Pauli paramagnetic effect. In the weak-coupling limit, the orbital limiting field can be quantified using the Werthamer-Helfand-Hohenberg (WHH) equation described as \cite{werthamer1966, helfand1966}:
\begin{equation}
    H_{C2}^{\mathrm{orb}}(0) = -\alpha T_C \left. \frac{dH_{C2}(T)}{dT}\right|_{T=T_{C}}
    \label{eqn6:Hc2_orb}
\end{equation}
Here, $\alpha$ is 0.693 for a dirty limit BCS superconductor (Section \ref{subsec4}). The orbital limiting field $H_{C2}^{\mathrm{orb}}(0)$ is found to be 1.23(2) T by using the slope value of $H_{C2}(T)$ near $T_C$ in Eq. \ref{eqn6:Hc2_orb}. The Pauli paramagnetic field for the weakly coupled BCS superconductors is given as $H_{C2}^{\mathrm{P}}(0) = 1.84 T_C$ \cite{chandrasekhar1962, clogston1962}. $H_{C2}^{\mathrm{P}}(0)$ is determined to be 8.52 T using $T_C=4.63$ K. The Maki parameter defined as $\alpha_\mathrm{M}=\sqrt{2} H_{c2}^{\mathrm{orb}}(0)/H_{c2}^{\mathrm{P}}(0)$ is used to calculate the relative strength of the orbital limiting effect to the Pauli paramagnetic effect \cite{maki1966}. The obtained value of $\alpha_\mathrm{M}$ is 0.20(4), suggesting a small contribution of the Pauli paramagnetic field to the pair breaking.

\textbf{Heat Capacity:} 
Heat capacity measurements were also carried out to analyze the thermal characteristics of Nb-Mo-Ru-Re-Ir under zero magnetic fields. The results are shown in the inset of \figref{Fig3}(a). A substantial increase in the heat capacity at $T_C$ of 4.54(6) K dictates the presence of bulk superconductivity in this HEA. This transition is consistent with magnetization and resistivity measurements. The Debye model provides the best fit for C(T)/T in the normal state:
\begin{equation}
    \frac{C(T)}{T} = \gamma_n +\beta_3T^2 +\beta_5T^4
    \label{eqn7:Debye}
\end{equation}
where $\gamma_n$ represents the Sommerfeld coefficient, corresponding to the electronic contribution, while $\beta_3$ and $\beta_5$ account for the phononic and anharmonic contributions, respectively. The extrapolation of the fit to 0 K yields $\gamma_n= 3.78(9)$ mJ-mol$^{-1}\mathrm{K}^{-2}$, $\beta_3 =0.042(5)$ mJ-mol$^{-1}\mathrm{K}^{-4}$, and $\beta_5=0.22(5)\ \mu$J-mol$^{-1}\mathrm{K}^{-6}$. The Debye temperature, $\theta_\mathrm{D}$ can be determined using $\beta_3$ according to the following equation \cite{Kittel}:
\begin{equation}
    \theta_\mathrm{D} = \left(\frac{12\pi^4RN}{5\beta_3}\right)^{\frac{1}{3}}
    \label{eqn8:thetaD}
\end{equation}
\begin{figure*}[ht]
\includegraphics[width=2.0\columnwidth, origin=b]{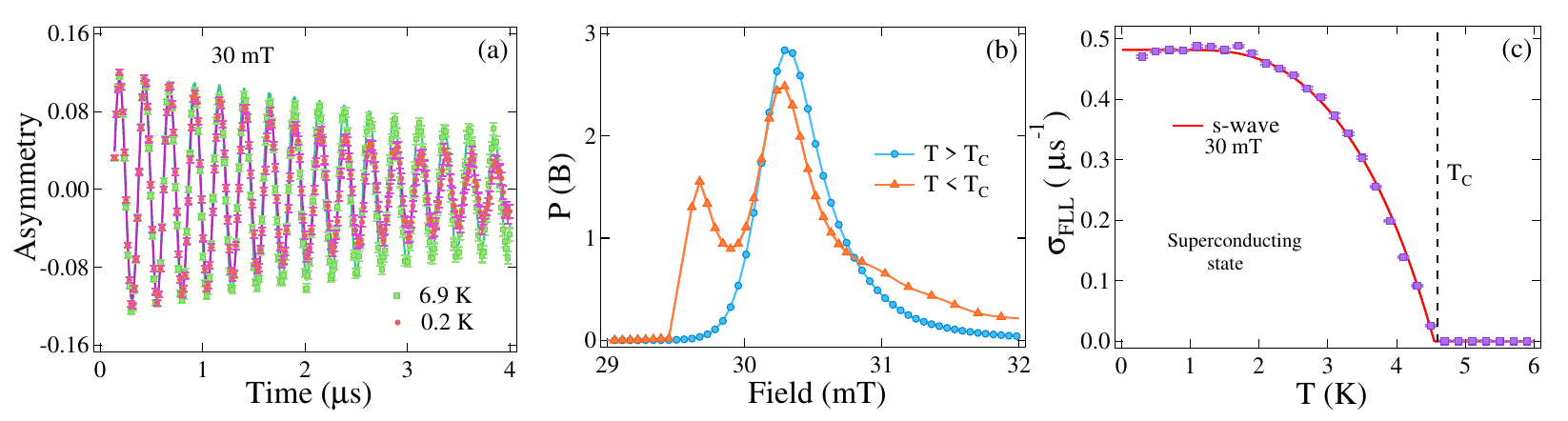}
\caption{\label{Fig4}(a) Transverse field (TF) $\mu SR$ spectra measured below (0.2 K) and above (6.9 K) $T_C$ in a 30 mT applied field. The solid lines are fits to the data using Eq. \ref{eqn14:TF asym}. (b) The probability distribution of magnetic field below and above $T_C$. (c) The temperature-dependent $\sigma_{\mathrm{FLL}}$ extracted from TF measurements and fitted with the BCS s-wave model (solid red line).}
\end{figure*}Here, $R=8.31\ \mathrm{Jmol}^{-1}\mathrm{K}^{-1}$ denotes the universal gas constant, and $N$ denotes the total number of atoms per formula unit, which is 1 for Nb-Mo-Ru-Re-Ir. After inserting the values of $R$, $N$, and $\beta_3$ in Eq. \ref{eqn8:thetaD}, $\theta_\mathrm{D}$ is estimated to be 359(42) K. The Sommerfeld coefficient in a non-interactive system is directly governed by the density of states at the Fermi level, $D_C(E_\mathrm{F})$, as follows:
\begin{equation}
    \gamma_n = \left(\frac{\pi^2k^2_\mathrm{B}}{3}\right)D_C(E_\mathrm{F})
    \label{eqn9:DOS}
\end{equation}
where $k_\mathrm{B}$ is the Boltzmann's constant and has a value of $1.38\times10^{-23}\ \mathrm{JK}^{-1}$. By replacing $\gamma_n= 3.78(9)$ mJ-mol$^{-1}\mathrm{K}^{-2}$, $D_C(E_\mathrm{F})$ is calculated to be 1.60(4) states/eV-f.u. The electron-phonon coupling constant $\lambda_{e-ph}$, proposed by McMillan, measures the strength of interaction between electrons and phonons and can be determined by the following expression using $\theta_\mathrm{D}$ and $T_C$ \cite{mcmillan1968transition}:
\begin{equation}
\lambda_{e-ph} = \frac{1.04+\mu^{*}\ln{\left(\theta_\mathrm{D}/1.45T_{C}\right)}}{\left(1-0.62\mu^{*}\right)\ln{\left(\theta_\mathrm{D}/1.45T_{C}\right)}-1.04}
\label{eqn10:e-ph}
\end{equation}
where the screened Coulomb potential $\mu^{*}$ is taken as 0.13 for inter-metallic compounds. $\lambda_{e-ph}$, is found to be 0.59(7), indicating weakly coupled superconductivity in Nb-Mo-Ru-Re-Ir.

The electronic heat capacity at low temperatures is evaluated by subtracting the lattice contribution from the total heat capacity ($C_{el}(T)=C_{tot}(T)-\beta_3T^3-\beta_5T^5$) which provides $\Delta C_{el}/\gamma_nT_c=1.18(5)$, slightly less than the BCS value of 1.43. The following s-wave model for fully gaped isotropic superconductors accurately describes the low-temperature behavior of the electronic heat capacity for the normalized entropy S, shown in \figref{Fig3}(a).
\begin{equation}
\frac{S}{\gamma_{n}T_{C}} = -\frac{6}{\pi^2}\left(\frac{\Delta(0)}{k_\mathrm{B}T_{C}}\right)\int_{0}^{\infty}\left\{f\ln{f}+(1-f)\ln{1-f}\right\}dy 
\label{eqn11:s-wave}
\end{equation}
where $f(\xi)=\left\{\exp\left(E(\xi)/k_\mathrm{B}T)\right)+1\right\}^{-1}$ represents the Fermi function, $E(\xi)=\sqrt{\xi^{2}+\Delta^{2}(t)}$ represents the normal electron energy relative to the Fermi level, and $y=\xi/\Delta(0)$, $t=T/T_{c}$ and $\Delta(t)=\tanh{\left\{1.82(1.018((1/t)-1)^{0.51}\right\}}$ represent an approximation of the BCS gap. The normalized entropy is related to the electronic heat capacity in the superconducting state by the following expression:
\begin{equation}
\frac{C_{el}}{\gamma_{n}T_{c}} = t\frac{d(S/\gamma_{n}T_{c})}{dt}
\label{eqn12:gap}
\end{equation}
Fitting the electronic heat capacity with Eq. \ref{eqn11:s-wave} provides the superconducting gap, $\Delta(0)/k_\mathrm{B}T_C=1.78(2)$ indicating weakly coupled BCS behavior, as it is close to the BCS value of 1.76 \cite{padamsee1973quasiparticle}.

Heat capacity measurements were also performed under various applied magnetic fields to gain deeper insight into the nature of the superconducting gap (shown in \figref{Fig3}(b)). Initially, the electronic heat capacity for each field was evaluated by removing the lattice contribution and fitted with the following equation in the superconducting state \cite{motla2022superconducting}:
\begin{equation}
    \frac{C}{T}=\gamma+\frac{A}{T}\exp\left(\frac{-bT_C}{T}\right)
    \label{eqn13:structure}
\end{equation}
where $A$ represents a quantity that depends on the material, and $b$ is the superconducting gap $\Delta(0)/k_\mathrm{B}T_C$. The coefficient $\gamma$ shows a linear increase with increasing applied field, as shown in \figref{Fig3}(c). The linear dependence of $\gamma$ on the applied field arises from a growing number of vortices with increasing field strength for an isotropic superconducting gap where quasiparticle excitations are confined inside the vortex cores \cite{isobe2016superconductivity, nakai2004field}.
This behavior suggests an isotropic superconducting gap, whereas $\gamma(H)$ has a square root dependence on the applied field if nodes are present.

\subsection{Muon Spin Relaxation and Rotation Measurement}

\textbf{Transverse field $\mu\text{SR}$:} The muon spin relaxation and rotation ($\mu$SR) technique was employed to explore the microscopic characteristics of the superconducting ground state \cite{hillier2022muon}. The superconducting gap symmetry of Nb-Mo-Ru-Re-Ir was thoroughly examined by transverse field (TF) $\mu$SR measurements. In this configuration, the magnetic field was oriented perpendicular to the initial muon spin. The sample was initially cooled through the superconducting transition in an applied magnetic field of 30 mT ($H_{C1}<H<H_{C2}$) to generate a well-ordered flux line lattice (FLL), with data acquisition performed during the subsequent warming process. 
\figref{Fig4}(a) shows the asymmetry spectra at 30 mT acquired below and above $T_C$. The rapid polarization relaxation observed below $T_C$ is ascribed to the inhomogeneous field distribution arising from the formation of a flux line lattice (FLL) in the mixed state. Conversely, the diminished depolarization above $T_C$ reflects the contribution of the magnetic field due to the randomly oriented static nuclear moments. \figref{Fig4}(b) represents the probability distribution of the magnetic field in the mixed state (orange curve) and the normal state (blue curve) obtained by the maximum entropy algorithm (MaxEnt). The TF asymmetry spectra are best described by a sinusoidal oscillating function, damped by a Gaussian relaxation envelope combined with an oscillatory background component \cite{weber1993magnetic, maisuradze2009comparison}: 
\begin{equation}
    \begin{split}
    A(t)&=A_0\exp(-\frac{1}{2}\sigma^2t^2)\cos{\left(\omega_1 t+\phi\right)} \\
    &+A_1\cos{\left(\omega_2t+\phi\right)}
    \end{split}
    \label{eqn14:TF asym}
\end{equation}
\begin{figure}[ht]
\includegraphics[width=0.9\columnwidth, origin=b]{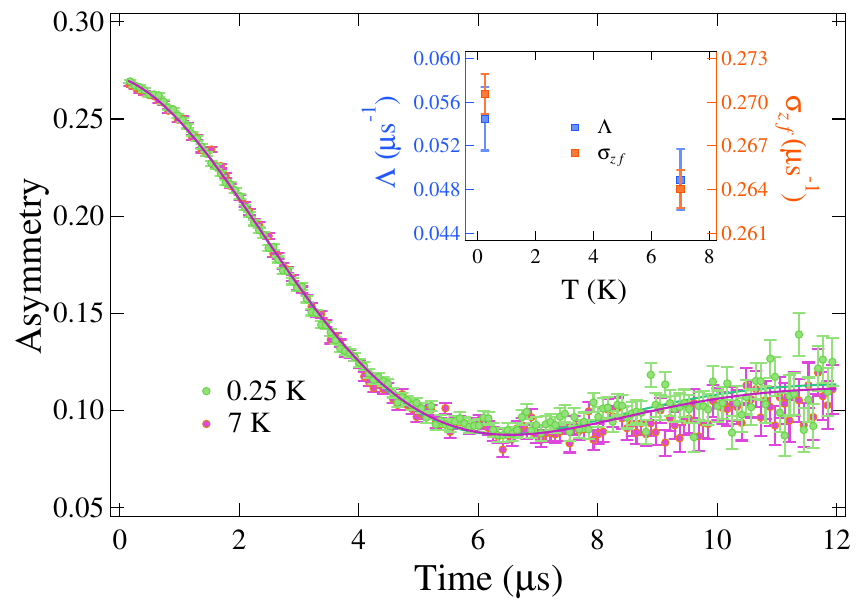}
\caption{\label{Fig5} Zero field (ZF) asymmetry spectra at 0.25 K (below $T_C$) and 7 K (above $T_C$) fitted with Eq. \ref{eqn17:ZF}. The inset shows the fitting parameters, $\sigma_{zf}$ and $\Lambda$ extracted below and above $T_C$.}
\end{figure}Here, the first term represents the signal originating from the sample, and the second term accounts for the background signal from the silver holder. The initial asymmetries and muon precession frequencies associated with the sample and background are represented by $A_0$ and $A_1$, and  $\omega_1$ and $\omega_2$, respectively, while $\sigma$ represents the depolarization rate. The depolarization rate $\sigma$ in Eq. \ref{eqn14:TF asym} consists of the contribution from nuclear dipole moments, $\sigma_N$, and the contribution from the FLL in the superconducting state, $\sigma_{FLL}$, and is expressed as $\sigma^2 = \sigma_N^2+\sigma_{FLL}^2$. The temperature dependence of $\sigma_{FLL}$ can be obtained by subtracting the temperature-independent depolarization of the nuclear moments $\sigma_N=0.3491(4)\ \mu\mathrm{s}^{-1}$ from the total depolarization rate, and is illustrated in \figref{Fig4}(c). At low temperatures, the increase in the relaxation rate can be attributed to enhanced vortex pinning and disorder, as discussed by Joel et. al. \cite{barker2018superconducting}. $\sigma_{FLL}$ is subsequently related to the penetration depth, $\lambda$ through the following equation \cite{brandt2003properties}:
\begin{equation}
    \frac{\sigma_{FLL}(T)}{\gamma_\mu^2} = \frac{0.00371\Phi_0^2}{\lambda^4(T)}
    \label{eqn15:sigma}
\end{equation}
where $\gamma_{\mu} = 2\pi\times135.5\ \mathrm{MHzT}^{-1}$ is the muon gyromagnetic ratio and $\Phi_0$ is the flux quanta. The temperature dependence of the penetration depth for the s-wave superconductor in the dirty limit can be described by the equation considering London's approximation \cite{carrington2003magnetic}:
\begin{equation}
    \frac{\sigma_{FLL}(T)}{\sigma_{FLL}(0)}=\frac{\lambda^{-2}(T)}{\lambda^{-2}(0)}=\frac{\Delta(T)}{\Delta(0)}\tanh{\left(\frac{\Delta(0)}{2k_\mathrm{B} T}\right)}
    \label{eqn16:Mu s-wave}
\end{equation}Here, $\Delta(t)=\tanh{\left\{1.82(1.018((1/t)-1)^{0.51}\right\}}$ represents the approximation of the temperature dependence of the energy gap for BCS superconductors. This function accurately describes the behavior of the superconducting depolarization rate where the solid red line represents the fit in \figref{Fig4}(c) and the extrapolated value of the superconducting gap is $\Delta(0)/k_\mathrm{B}T_C=2.34(5)$. This value is slightly larger than that obtained from the heat capacity measurements.This small discrepancy has also been observed in Re$_{5.5}$Ta \cite{arushi2020unconventional} and La$_7$Rh$_3$ \cite{singh2020time}.
The high value of the superconducting gap relative to the BCS value indicates the presence of the strong-coupling nature of the electron pairs in Nb-Mo-Ru-Re-Ir. The penetration depth at 0 K from the $\mu$ SR measurement, $\lambda^{\mu}(0)$, was evaluated using Eq. \ref{eqn15:sigma} resulting in $\lambda^{\mu}(0) = 396(8) \ \mathrm{nm}$. This value is somewhat higher than that extracted from the magnetization data. A similar observation has been made in the HEA superconductor Nb-Re-Zr-Hf-Ti \cite{motla2022superconducting}.

\textbf{Zero field $\mu\text{SR}$:} To investigate the potential existence of any spontaneous magnetization or time-reversal symmetry (TRS) breaking in the superconducting state of Nb-Mo-Ru-Re-Ir, zero field (ZF) $\mu\text{SR}$ measurements were carried out, as muons exhibit high sensitivity to small magnetic fields. The time evolution of the ZF asymmetry spectra acquired both above and below $T_C$ are shown in \figref{Fig5}. An increase in the relaxation rate in the superconducting state would signal the presence of spontaneous magnetization, thereby indicating broken TRS \cite{ghosh2020recent}. In the absence of any electronic magnetic moments, the depolarization of the muon is solely attributed to the effects of randomly oriented nuclear moments. The static Kubo-Toyabe times exponential decay function, plus a flat background component, shown below, provides a good fit for both spectra \cite{hayano1979zero}.
\begin{equation}
  \begin{split}
    A(t)&=A_0\left[\frac{1}{3}+\frac{2}{3}(1-\sigma^2_{zf}t^2)\exp(-\frac{1}{2}\sigma^2_{zf}t^2)\right]\exp{-\Lambda t} \\
    &+ A_{bg}
    \label{eqn17:ZF}
  \end{split}
\end{equation}Here, $\sigma_{zf}$ corresponds to the muon spin relaxation rate, which originates from the local fields generated by static and randomly oriented nuclear moments. $\Lambda$ corresponds to the electronic relaxation rate, while $A_0$ and $A_{bg}$ represent the initial asymmetry contributions from the sample and the background, respectively. The difference in the fit parameters, as shown in the inset of \figref{Fig5}, was found to be $\Delta\sigma_{zf} = 0.004\ \mu\mathrm{s}^{-1}$ and $\Delta\Lambda = 0.005\ \mu\mathrm{s}^{-1}$. This insignificant difference rules out the possibility of any spontaneous magnetization, thereby indicating that time-reversal symmetry is preserved in Nb-Mo-Ru-Re-Ir.
\subsection{Electronic Properties and Uemura Plot}
\label{subsec4}
A set of experimentally determined parameters, including the carrier concentration ($n$), residual resistivity ($\rho_0$) and the Sommerfeld coefficient ($\gamma_n$), are used to evaluate electronic parameters and to gain a deeper understanding of the experimental results for Nb-Mo-Ru-Re-Ir. The effective mass ($m^*$) can be determined from $\gamma_{n}$ and $n$ using $\gamma_{n} = \left(\frac{\pi}{3}\right)^{2/3}\frac{k_\mathrm{B}^{2}m^{*}n^{1/3}}{\hbar^{2}}$, where $\hbar=1.05\times10^{-34}$ Js represents the reduced Planck constant. By substituting the values of $\gamma_n=3.78(9)$ mJ-mol$^{-1}\mathrm{K}^{-2}$ and $n=2.4(5)\times10^{28}\ \mathrm{m}^{-3}$ obtained from Hall measurements, the effective mass $m^*$ is determined to be $9.0(9)\ m_e$. The Fermi velocity $v_\mathrm{F}$, is determined to be $v_\mathrm{F}=1.1(2)\times10^5 \ \mathrm{ms}^{-1}$ using the relation $n=\frac{1}{3\pi^2}\left(\frac{m^*v_\mathrm{F}}{\hbar}\right)^3$. The mean free path ($l$) can then be calculated using the formula $l = \frac{3\pi^2\hbar^3}{e^2\rho_0m^{*2}v^2_{\mathrm{F}}}$. Substituting the values of $\rho_0=128.51(5)\ \mu\Omega$-cm, $m^*$, and $v_\mathrm{F}$ gives $l=11.8(8)$ \AA. The BCS coherence length ($\xi_0$) is related to $v_{\mathrm{F}}$ and $T_C$ by the expression $\xi_0=\frac{0.18\hbar v_{\mathrm{F}}}{k_\mathrm{B}T_C}$, which results in $\xi_0=339(63)$ \AA. \begin{figure}[ht]
\includegraphics[width=0.9\columnwidth, origin=b]{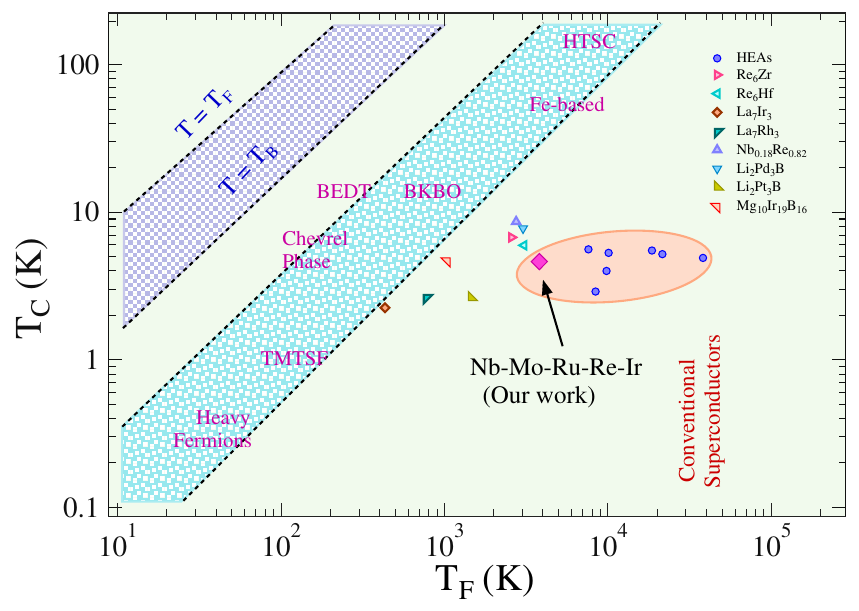}
\caption{\label{Fig6} Classification of superconductors using the Uemura plot represented by $T_C$ versus $T_\mathrm{F}$. The blue-colored region represents the unconventional superconductors. The pink-colored square represents Nb-Mo-Ru-Re-Ir.}
\end{figure}
Superconductors are classified as clean or dirty limit based on the ratio of coherence length to mean free path ($\xi_0/l$). A high ratio, $\xi_0/l$ >> $1$, for Nb-M-Ru-Re-Ir, indicates a dirty limit superconductivity. 

Uemura \textit{et al.} proposed a classification scheme for superconductors using the ratio of superconducting transition temperature ($T_C$) to the Fermi temperature ($T_\mathrm{F}$) \cite{uemura1988systematic, uemura1989universal, uemura1991basic}. Superconductors with $0.1\geq T_C/T_\mathrm{F}\geq0.01$ are classified as unconventional superconductors. In contrast, superconductors with $T_c/T_\mathrm{F}\leq 0.01$ are classified as conventional superconductors. The Fermi temperature of a three-dimensional non-interacting system can be given as \cite{hillier1997classification}:
\begin{equation}
    k_\mathrm{B}T_\mathrm{F} = \frac{\hbar^2}{2m^*}\left(3\pi^2n\right)^{\frac{2}{3}}
    \label{eqn18:Fermitemp}
\end{equation}
After inserting the values of $m^*$ and $n$, Eq. \ref{eqn18:Fermitemp} gives $T_\mathrm{F}=3.8(9)\times 10^3$ K. The ratio $T_C/T_\mathrm{F}$ is found to be 0.0011(2), which places it far from the unconventional band but close to other unconventional superconductors such as Re$_6$Zr in Fig.\ref{Fig6}. Table \ref{Tab2} summarizes all normal and superconducting parameters.

\section{CONCLUSION}

In summary, we have successfully synthesized an equiatomic high-entropy alloy, Nb-Mo-Ru-Re-Ir, which crystallizes in an hcp structure consistent with the Hume-Rothery rule. A comprehensive investigation of its normal and superconducting states was conducted through magnetization, electrical resistivity, heat capacity, and $\mu$SR measurements. Our analysis confirms the presence of bulk superconductivity, with a $T_C$ of 4.63(2) K. The observed upper critical field from all the measurements is close to the Pauli paramagnetic limit, indicating the presence of some unconventional behavior. The heat capacity and TF $\mu$ SR measurements suggest fully gapped isotropic superconductivity. ZF $\mu$SR measurements confirm that the time-reversal symmetry is preserved in the superconducting state of Nb-Mo-Ru-Re-Ir. 
\begin{table}[ht]
\caption{Parameters in the superconducting and normal state of Nb-Mo-Ru-Re-Ir.}
\label{Tab2}
\begin{center}
\begingroup
\setlength{\tabcolsep}{8
pt}
\begin{tabular}{c c c} 
\hline\hline
Parameters & Unit & Value  \\ [1ex]
\hline
$T_{C}$& K& 4.63(2)\\ 
VEC& & 7\\             
$H_{C1}(0)$& mT& 13.59(9)\\                       
$H_{C2}^{Mag}(0)$& T& 8.0(2)\\
$H_{C2}^\mathrm{P}(0)$& T&8.52\\
$H_{C2}^{Orb}(0)$& T& 5.7(5) \\
$\xi_\mathrm{GL}(0)$& nm & 6.4(1)\\
$\lambda_\mathrm{GL}^{Mag, \mu}(0)$& nm & 208(9), 396(8)\\
$k_\mathrm{GL}$& &32.5(6)\\
$\gamma_{n}$&  mJ-mol$^{-1}$K$^{-2}$& 3.78(2) \\
$\theta_\mathrm{D}$& K& 359(42)\\
$\lambda_{e-ph}$& &0.59(7)\\
$\Delta C_{el}/\gamma_nT_C$ & &1.18(5)\\
$\Delta^{HC, \mu}(0)/k_\mathrm{B}T_C$& &1.78(2), 2.34(5)\\
$\xi_{0}/l_{e}$& &  28(7)\\
$v_{\mathrm{F}}$& 10$^{5}$ ms$^{-1}$& 1.1(2)\\
$n$& 10$^{28}$m$^{-3}$& 2.4(5)\\
$T_{\mathrm{F}}$&10$^{3}$ K& 3.8(9)\\
$m^{*}$/$m_{e}$&  & 9.0(9)\\
[1ex]
\hline\hline
\end{tabular}
\endgroup
\end{center}
\end{table}
These results provide insights into the superconducting pairing mechanism within disordered high-entropy alloys and serve as a crucial data point for exploring the impact of crystal structure, spin-orbit coupling, disorder, and hardness on superconducting ground states, ultimately facilitating the discovery of novel superconducting high-entropy alloys.

\section{ACKNOWLEDGEMENT} 
R.P.S. acknowledges the SERB Government of India for the Core Research Grant No. CRG/2023/000817. We thank ISIS, STFC, U.K. for providing the beamtime for the $\mu$SR experiments.
\pagestyle{plain}
\addcontentsline{toc}{chapter}{Bibliography}

\end{document}